\documentclass[aps,prl,reprint,floatfix,amsmath,amssymb,
superscriptaddress]{revtex4-2}

\usepackage{graphicx}
\usepackage{epstopdf}
\usepackage[dvipsnames]{xcolor}
\usepackage{hyperref}
\usepackage{bm}
\usepackage{printlen}
\usepackage{units}
\usepackage{braket}

\hypersetup{
        colorlinks,
   citecolor=blue,
   filecolor=blue,
   linkcolor=blue,
   urlcolor=blue,
   breaklinks=true}

\renewcommand{\vec}[1]{\mathbf{#1}}

\hypersetup{hidelinks}

\begin{document}

\title{Engineering SYK interactions in disordered graphene flakes \\
under realistic experimental conditions}

\author{Marta  Brzezi\'{n}ska}

\thanks{These two authors contributed equally.}

\affiliation{Institute of Physics, {\'E}cole Polytechnique F{\'e}d{\'e}rale de Lausanne,  Lausanne, CH 1015, Switzerland}

\author{Yifei Guan}

\thanks{These two authors contributed equally.}

\affiliation{Institute of Physics, {\'E}cole Polytechnique F{\'e}d{\'e}rale de Lausanne,  Lausanne, CH 1015, Switzerland}

\author{Oleg V. Yazyev}

\affiliation{Institute of Physics, {\'E}cole Polytechnique F{\'e}d{\'e}rale de Lausanne,  Lausanne, CH 1015, Switzerland}

\author{Subir Sachdev}

\affiliation{Department of Physics, Harvard University, Cambridge, Massachusetts 02138, USA}

\author{Alexander Kruchkov}

\affiliation{Institute of Physics, {\'E}cole Polytechnique F{\'e}d{\'e}rale de Lausanne,  Lausanne, CH 1015, Switzerland}

\affiliation{Department of Physics, Harvard University, Cambridge, Massachusetts 02138, USA}  

\affiliation{Branco Weiss Society in Science, ETH Zurich, Zurich, CH 8092, Switzerland}


\begin{abstract}
We model  SYK (Sachdev-Ye-Kitaev) interactions in disordered graphene flakes up to 300 000 atoms ($\sim$100 nm in diameter) subjected to an out-of-plane magnetic field $\vec B$ of 5-20 Tesla within the tight-binding formalism. We investigate two sources of disorder: (i) irregularities at the system boundaries, and (ii) bulk vacancies,---for a combination of which we find  conditions which could be favorable for the formation of the phase with SYK features under realistic experimental conditions  above the liquid helium temperature.
\end{abstract}

\maketitle

There has been significant recent interest in the condensed matter community of a holographic gravitational description of correlated electron 
systems \cite{Hartnoll2018,Hartnoll2009}.
A  model in this direction is the Sachdev-Ye-Kitaev (SYK) model \cite{Sachdev1993, Kitaev2015, Kitaev2015}, which describes from the condensed matter perspective a set of $N$ electrons in a dispersionless quantum state (a flat band), interacting strongly yet randomly all-to-all,
\begin{equation}
\mathcal H_{\text{SYK}} = \sum_{ijkl}^{N} J_{ijkl} \, c^{\dag}_i c^{\dag}_j c^{}_k  c^{}_l .
\label{eq:SYK} 
\end{equation}
Here $c^{\dag}_i$ ($c^{}_i$) are fermionic creation (annihilation) operators, and $J_{ijkl}$ are random couplings in all indices (the model works beyond the Gaussian randomness \cite{Krajewski2019}). Despite its attractive mathematical properties such as exact solvability in the large $N$ limit with nearly conformal properties~\cite{Maldacena2016,Gu2020}, mapping on the Jackiw-Teitelboim gravity \cite{Kitaev2018}), and importance for condensed matter physics (including strange metallicity~\cite{Shenoy2018, Cha2020, CGPS} and superconductivity on the basis of SYK model \cite{Wang2020c, Classen2021, Inkof2022}),  a direct experimental realization is currently missing.

A theoretical proposal for a potential experimental platform for the electronic SYK model, given in Ref. \cite{Chen2018}, is a graphene dot with irregular boundaries placed in an external magnetic field. Ref. \cite{Chen2018} studied  $\sim$2000 atoms (5 nm in radius) in a field of $\sim$3200 T. 
However, the magnetic fields employed in Ref. \cite{Chen2018} exceed capabilities within the laboratory realm. 
Modern condensed matter facilities operate with quantum transport  at magnetic fields up to 16-20 T, and the highest accessible magnetic fields in DC operation are of 45 T \cite{Hahn2019}. At the same time, graphene preparation and chemical etching procedures pose limits on the controllable size and shape of a flake, allowing flexible operational capabilities with the flakes size of hundred nanometers and above, but not for a few nanometers size. In this regard, a great challenge is to overcome these obstacles to engineer a realistic graphene flake, which could host relevant interactions in the experimentally accessible magnetic fields of 5-20 T.

 \begin{figure*}[t]
\includegraphics[width= 0.95 \textwidth]{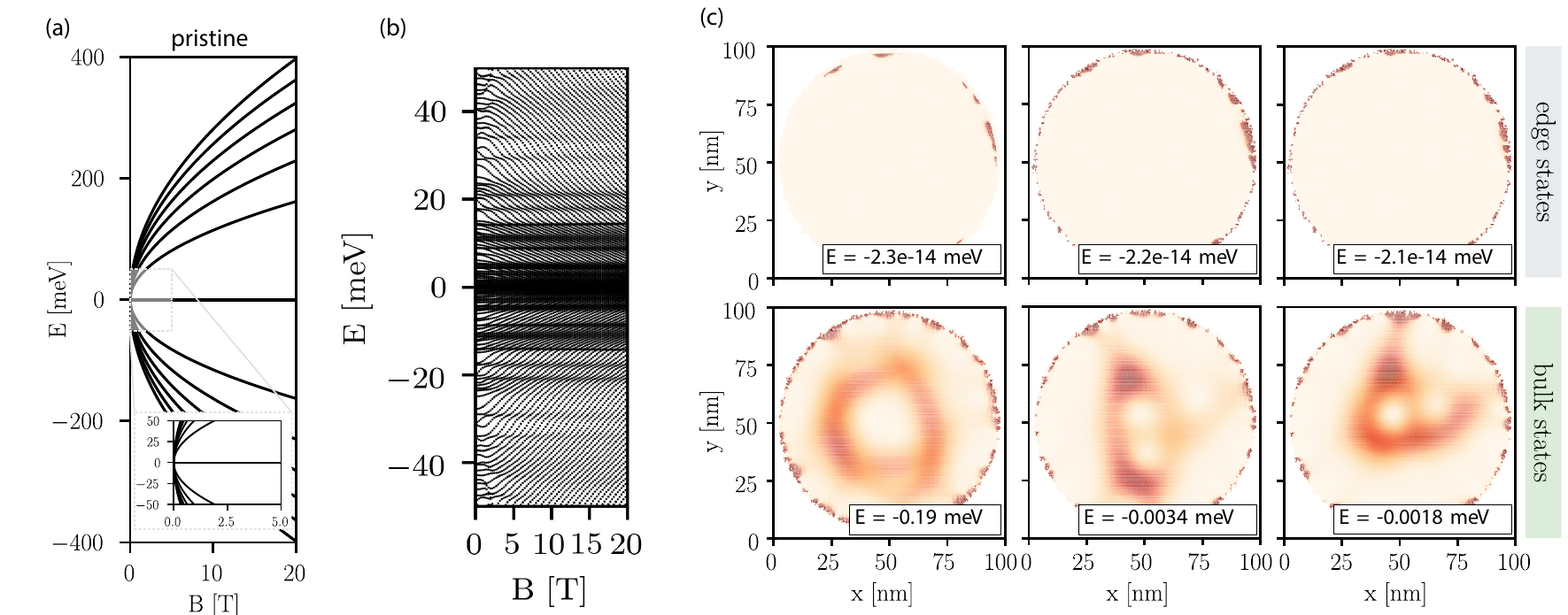}
\caption{\textbf{Bulk  and edge states in disordered graphene flake of 100 nm in strong magnetic fields}.  (a) Energy spectrum as a function of magnetic field, expected in pristine graphen; inset shows the same spectrum in energy range [-50 meV; + 50 meV] to compare with panel (b); (b) De-facto energy spectrum, observed in a strongly disordered graphene flake of size 100 nm; higher Landau levels ($|n|$$>$$0$) are not recognized in the energy range [-50,+50 meV]; (c) probability densities $|\psi|^2$ of exemplary bulk and edge states.  Randomly localized bulk states shown in bottom of (c) are the building blocks for constructing the SYK-like interactions (Fig. \ref{Fig2}). }
\label{Fig1}
\end{figure*}

In this paper we report large-scale calculations on large graphene flakes involving up to 300 000 carbon atoms (corresponding to flake size $\approx$ 100 nm) placed under realistic experimental conditions. We find that upon choosing a well-disordered flake, we can reach favorable experimental conditions with SYK strength $J$$\sim$$45$ meV (we use standard normalization $J^2 = {2 N^3} \langle |J_{ijkl}|^2 \rangle $), and a mesoscopic number of SYK fermions $N$, typically around 40 in our calculations in the magnetic fields of 10-20 T. 
We further model: (i) the role of chemical etching \cite{Wang2013}, or local anodic oxidation with an atomic force microscopy (AFM) tip \cite{Puddy2011, Masubuchi2011, Li2018}, by varying the shape of the flake and the size of edge disorder.  (ii) the role of bulk vacancies created, for example, by focused ion beam (FIB) patterning \cite{Archanjo2012} or  hydrogen plasma treatment \cite{Despiau2016}. Our results speak in favor of formation of SYK-like interactions in the realistic range of parameters. In particular, we point out how the relative effect of melonic diagrams can be enforced by controlling the atomic vacancies concentration in the bulk.
By comparing the relevant energy scales~\cite{Kruchkov2020}, namely $t^2/J$ and $J/N$, we come to the conclusion that the engineered system has a set  of parameters where it could realize the SYK phase in the vicinity of the liquid helium operational temperatures, accessible magnetic fields, and suitable graphene flake scales.

\textit{Setup.} To construct the SYK model, one needs to employ the dispersionless quantum states (flat bands), in particular electronic states with nontrivial Bloch topology. 
Such flat band states have been classified in Ref.~\cite{Kruchkov2022}. The simplest of this construction are the Landau levels, which are characterized by Chern number $|C|$$=$$1$. 
In principle, one can use numerous 2D materials for this purpose, however we here limit ourself to the case of graphene \cite{Chen2018} for two reasons: (i) graphene monolayer is an intuitively understood system from both analytical and numerical viewpoint; (ii) there are existing experimental platforms satisfying criteria for this direction~\cite{Anderson2022}. 

\begin{figure}[t]
\includegraphics[width= 0.5 \textwidth]{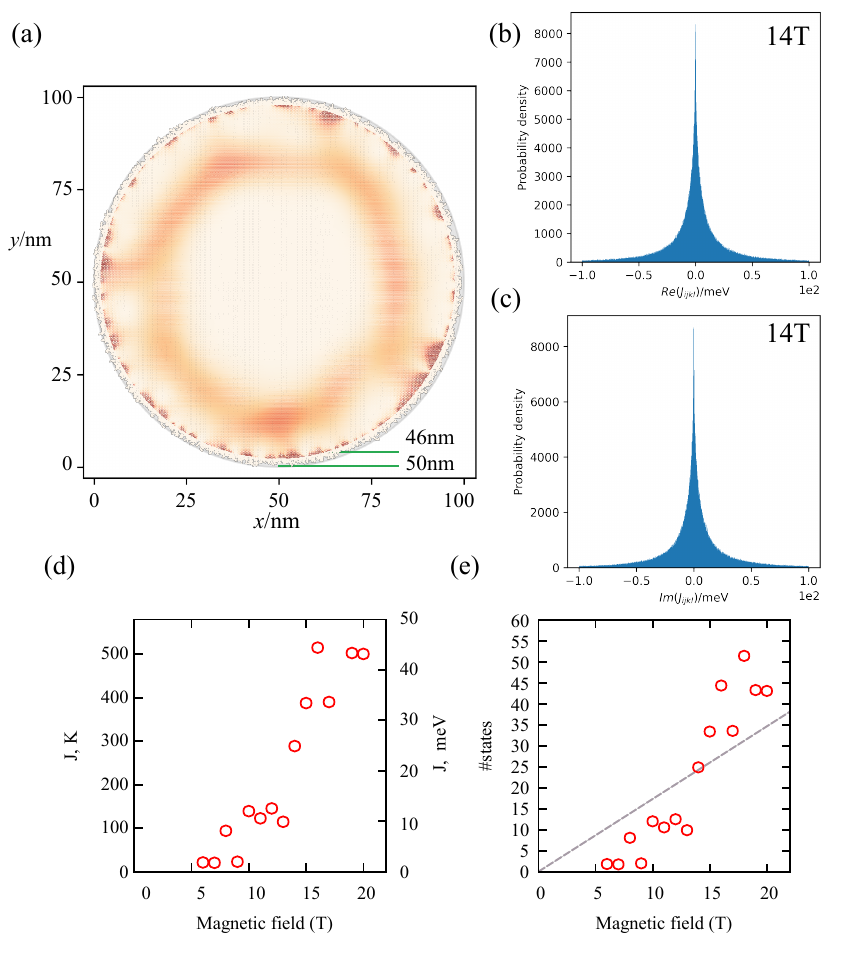}
\caption{\textbf{SYK-like interactions in the graphene flake of size 100 nm} (268'510 Carbon atoms) under realistic magnetic fields 5-20 T.  (a) The geometry of the flake used for numerical modeling, together with the visualization of a typical bulk state near the Fermi level.  (b-c) Distribution of real and imaginary parts $J_{ijkl}$ terms computed from Eq. \eqref{Jterms} with $\text{Re} \langle J_{ijkl} \rangle \approx 0$ and $\text{Im} \langle J_{ijkl} \rangle \approx 0$.  (d) The value of SYK interaction $J$, determined by the second moments in Eq. \eqref{eq:JmeV}. For calculation, we take bulk states distributed between [-1 meV, +1 meV] around the Fermi level. Panel (e) shows number of bulk states in the range $[-1$ meV,$+1$ meV] involved into $J_{ijkl}$ calculation; gray dashed line depicts ideal LLL case Eq.\eqref{eq:N0}.  }
\label{Fig2}
\end{figure}

Before proceeding to disordered graphene flakes, let us recall the physics of pristine (homogeneous and boundless) graphene in low-energy approximation. Upon application of out-of-plane magnetic field $B$, the electronic spectrum of pristine graphene is given by~\cite{Gusynin2005}
\begin{equation}
E_n = \pm v_F \sqrt{2 \hbar e B |n|},  
\label{eq:spectrum}
\end{equation}
where $e$ is electronic charge and $v_F$$\approx$$10^6$  m/s is the Fermi velocity. The lowest Landau level (LLL) is characterized by zero modes in the bulk ($n$$=$$0$). In the presence of chiral symmetry, the Aharonov-Casher argument~\cite{Aharonov1979} sets the number of electronic states in LLL as
\begin{equation}
N_0 = \frac{BA}{\Phi_0} , 
\label{eq:N0}
\end{equation}
where $A$ is the flake area, and $\Phi_0 = h/e =4.136\cdot 10^{-15}$ Wb is the magnetic flux quantum.  
 $N_0$ in Eq.~\eqref{eq:N0} sets the order of magnitude for the number $N$ of SYK states in the Hamiltonian in Eq.~\eqref{eq:SYK}, however we  see that $N$ within bandwidth $t=2$ meV around the Fermi level $N$ is fluctuating around $N_0$ due to strong  disorder effects in considered graphene flakes (see Fig. \ref{Fig2}). Still there is a certain qualitative similarity with the ideal LLL case, even that the flake is strongly disordered.

The typical electronic spectrum of a strongly disordered graphene flake is illustrated in Fig. \ref{Fig1}.  The flake has a disorder-free inner region of radius $R_1$, followed by disordered edge up to radius $R_2$ (46 nm and 50 nm in Figs. \ref{Fig1}, \ref{Fig2}).   Tight-binding (TB) calculations are performed with the conventional graphene model in magnetic fields \cite{Goerbig2010}, taking into account nearest-neighbors hoppings with Peierls substitution \footnote{For computational advance, we combine KWANT toolbox \cite{Groth2014} with FEAST algorithms \cite{Polizzi2009}}.
Figure \ref{Fig1} shows the results obtained from diagonalization of the TB model describing the 100 nm flake   consisting of around 270 000 of atoms. The first observation is that the electronic spectrum of a realistic disordered flake in the relevant energy range deviates significantly from its pristine counterpart, given by Eq. \eqref{eq:spectrum}: in all the realistic magnetic fields 0 to 20 T, we no longer observe the square-root behavior of eigenenergies as expected for pristine graphene; instead, the spectrum acquires a quantum-dot-like distribution \footnote{We note however that in much stronger magnetic flux $\sim h/e$, the well-defined structure from Eq. \eqref{eq:spectrum} reappears.}. 

We  distinguish the bulk states from the edge states  through the analysis of their localization properties. For this, we integrate  $|\Psi|^2$ within radius $R_1 + \delta$ ($\delta$$\to$$0$) \footnote{For numerical purposes, we use $\delta = 0.1 (R_2 - R_1)$ in the calculations in the main text.}. Due to the presence of irregularies on the boundaries, all the states are showing certain localization at the edge; however bulk stats have significant weight in bulk. If at least $50 \%$ of weight is localized in bulk, we label this state as a bulk state (In the Supplementary Material, we show that a different criterion, based on analysis of inversed participation ratio (IPR), leads to qualitatively similar results). 
While, as expected, the edge states are strongly localized at the irregular boundaries, the bulk states sway over all the flake diameter, being spread on the length scale of $\sim$100 nm.   
With such large length scale over which the bulk states are spread, the notion of distance is lost, and these states are interacting  randomly all-to-all, in the SYK spirit.

\textit{Calculation of the SYK terms.} With the bulk states randomly localized, and the kinetic energy quenched to $t<2$ meV, we construct the SYK-like states by introducing the Coulomb interaction in the basis of randomized bulk states~\cite{Chen2018, Wei2021}. We compute Sachdev-Ye-Kitaev interaction terms through
\begin{equation}
J_{ijkl} = \frac{1}{2}  
\sum_{\vec r_1} \sum_{\vec r_2}
 \Psi^*_i (\vec r_1) \Psi^*_j (\vec r_2) 
 U (\vec r_1 - \vec r_2) 
\Psi^{}_k (\vec r_1) \Psi^{}_l (\vec r_2) 
\label{Jterms}
\end{equation}
where $U(\vec r)$ is the screened Coulomb potential. Our results stand for different forms of the screened Coulomb potentials. To be specific, we adopt the values of renormalized interaction potentials for graphene as it was reported in Refs.~\cite{Wehling2011}. In particular, we adopt $U_{\text{NN}}=5.5$ eV, $U_{\text{NNN}}=4.1$ eV, $U_{\text{NNNN}}=3.6$ eV,  where (N)NN stands for (next)-nearest neighbor interactions.  
In calculation of $J_{ijkl}$ terms \eqref{Jterms}, we take only bulk states within the energy range $[-1$ meV, $+1$ meV] around the neutrality. This qualitatively corresponds to focusing on the states associated with what used to be LLL (see dashed line in Fig.\ref{Fig2}e), similar to Ref.~\cite{Chen2018}. For medium size flakes, we check that changing the bulk states range $[-1$ meV,$+1$ meV] to $[-5$ meV,$+5$ meV] does not change significantly the results for $J_{ijkl}$ calculations, since most of the bulk states are situated near the zero energy, and higher excited bulk states contribute marginally to SYK interactions $J_{ijkl}$. Hence, in what follows we proceed with bulk states within $[-1$ meV,$+1$ meV].

The key results for the large flake are summarized in Fig.~\ref{Fig2}. 
The statistical distribution of the complex-valued $J_{ijkl}$ terms is illustrated in Fig.~\ref{Fig2}(bc). The mean is zero, (Re $\langle J_{ijkl} \rangle \approx 0$, Im $\langle J_{ijkl} \rangle \approx 0$), which indicates that real and imaginary parts of $J_{ijkl}$ are independent. The overall distribution of the absolute values of $J_{ijkl}$ (in Fig.~\ref{Fig2}(c)) is quasirandom, but non-Gaussian as in conventional SYK models \cite{Gu2020,Maldacena2016}; however, this is not the problem for constructing SYK-like models \cite{Krajewski2019}.
We introduce the real-valued strength of SYK interactions $J$  as with normalization from counting melonic diagrams  \cite{Gu2020}
\begin{equation}
J =  \sqrt{2}  N^{3/2} \braket{J^{}_{ijkl} J_{ijkl}^*}^{1/2}. 
\label{eq:JmeV}
\end{equation}
We operate this quantity in meV and Kelvin for practical convenience. The results for the large flake are encouraging, with extracted $J$ of around 35 meV at 15 T (see Fig. \ref{Fig2}(d)). The number of SYK fermions peaks to 50-60 and we typically take around 40 of them for our calculations.

 \begin{figure}[b]
\includegraphics[width=1 \columnwidth]{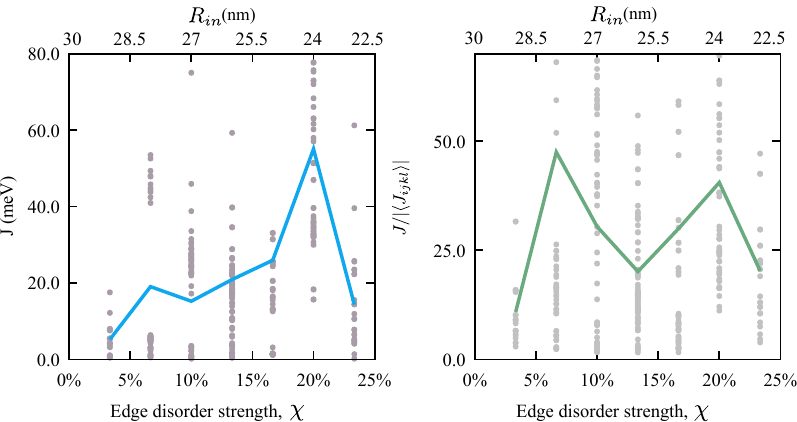}
\caption{\textbf{Enhancing SYK interactions through edge disorder $\chi$.}  (a) SYK coupling strength $J$, (b) dimensionless ratio $J/| \langle J_{ijkl} \rangle | $  as a function of edge disorder scale $\chi$ for the medium size flake at $B = 20$ T (approx. 100 000 atoms, 60 nm in diameter).  Solid lines correspond to the average over up to 20 disorder realizations (marked with grey points, some outside of plot range).  }
\label{Fig3}
\end{figure}

\begin{figure*}[t]
\includegraphics[width= 0.9 \textwidth]{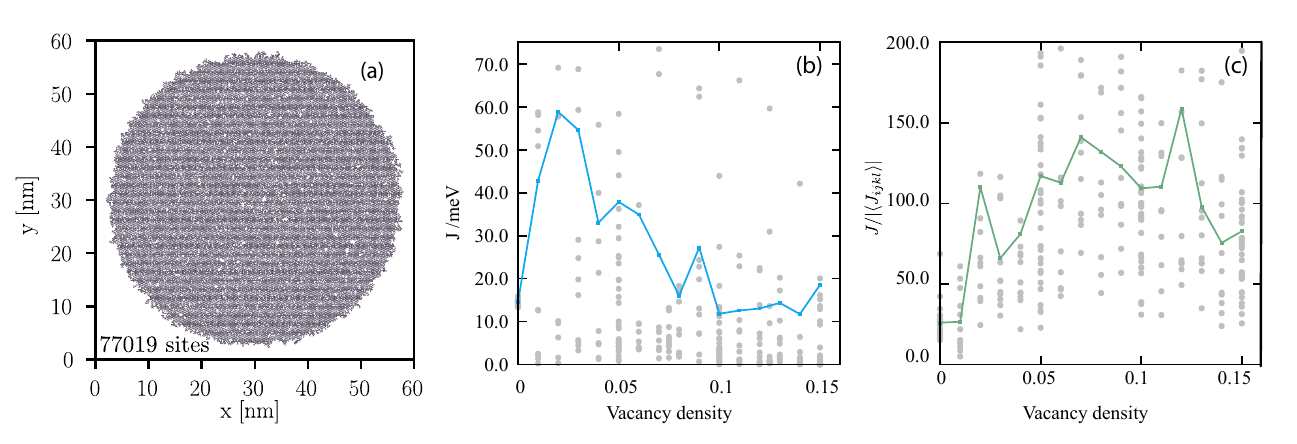}
\caption{\textbf{Tuning SYK strength $J$ with vacancy patterning.}   (a) illustration of the graphene flake with bulk atomic vacancies (approx. 77 000 atoms); (b) SYK strength $J$ and (c)   $J/| \langle J_{ijkl} \rangle | $ ratio as a function of the vacancies concentration  in the range from 0 \% to 15 \% of vacancies. Solid blue lines correspond to arithmetic average over 20 disorder realizations in the medium size flake with $R_{1} = 26$ nm and  $R_{2} = 30$ nm at $B = 20$ T (some data points (grey) are outside of the plot range, see SM for details). At 2 \% of vacancy concentration, we observe a clear maximum of SYK strength $J \sim 60$ meV. Above 15 \%,  the long-range order is destroyed for some disorder realizations and around 30 \%, all the systems are close to the percolation threshold for hexagonal lattice $p \approx 0.7$ \cite{Suding1999,Feng2008}.}
\label{Fig4}
\end{figure*}

\textit{Dependence on edge disorder scale}. We next address the question of how the edge disorder  $\chi = (R_2-R_1)/R_2$ influences the SYK interaction strength $J$.  This question is vital for the experiments, where only a limited number of methods  is available for shaping the flake of the size of 100-200 nm (chemical etching, FIB, hydrogen  plasma treatment). 
To optimize the numerical costs, we now turn to the medium size flakes of diameter 60 nm  
($\sim$ 100 000 atoms); these results are rescalable towards large graphene flake of size 300 000 atoms and more as in prototypes \cite{Anderson2022}.
We observe that the strength of the SYK interaction $J$ can be tuned by increasing the edge disorder $\chi$ (see Fig. \ref{Fig3}).  To quantify this effect we perform disorder averaging over dozens of flakes (Fig \ref{Fig3} uses up to 20 flake realizations). Typically $J$ scatters  from 5-10 meV  to nearly 60 meV upon increasing $\chi$ from $0$ to $20 \%$,  but some samples may  exhibit even larger values of $J$ above 80 meV (outside of plot range in Fig. \ref{Fig3}) in the intermediate regime.
Comparing with Fig. \ref{Fig2}, we come to the conclusion that the energy scale  $J \sim$ 20-40 meV is the most robust for the experiments. 
Moreover, the ratio of  $J/| \langle J_{ijkl} \rangle | $ in  Fig. \ref{Fig3}(b), which qualitatively points to the dominance of the melonic diagrams in the large-$N$ limit, increases under edge disorder.

\textit{Dependence on bulk vacancies and their concentration}. Another mechanism of control is behavior vacancies implanting as in Fig~\ref{Fig4}(a), performed e.g. via FIB tools.  In this case, the sample patterning can provide additional tuning knob to improve the properties towards SYK-like behavior. We here perform the calculations for the medium size flakes of 60 nm in $B = 20$ T. The results are shown in Fig.~\ref{Fig4}.  We start with the flake which has only moderate (non-optimized) edge disorder, reflected in $J \approx 15$ meV,  and gradually increase the number atomic vacancies. Fig~\ref{Fig4}(b) gives the dependence of the SYK interaction $J$ versus vacancy concentration in the bulk.  The first effect of vacancies patterning is a peak around $2 \%$ concentration with $J \approx 60$ meV and then the decrease of SYK coupling $J$ back to 10-20 meV at 15 \% concentration. The number of bulk states slowly grows with a vacancy concentration (see SM); we check that the bulk states are not exponentially localized on the atomic vacancies.  In Fig.~\ref{Fig4}(c), the ratio $J/| \langle J_{ijkl} \rangle | $ is  improved by adding a moderate amount of vacancies (at 0 \% this ratio is $J/| \langle J_{ijkl} \rangle | \approx 25 $, at 2 \% it is $J/| \langle J_{ijkl} \rangle | \approx 110$), hence  promoting the role of melonic diagrams \cite{Gu2020}.  Therefore, we come to conclusion that even a modest vacancy concentration of $2 \%$ improves the properties of the SYK flake.  For the flakes of size 100 nm, we recommend removing $\sim 6000$ Carbon atoms.

\textit{Discussion of the temperature scales}. Finally,  we perform the analysis of the relevant energy scales. The key energy scale is 
$J \approx 35  \, \text{meV}  $ (taken at experimentally relevant $B = $16 T from Fig. 2). The SYK model is well-defined in the region $T \ll J$. 
Furthermore, there are lower bounds on the temperature  coming from (i) finite bandwidth of the flat band, and (ii) mesoscopic effects in the SYK Hamiltonians. Both these temperatures scales appear in quantum transport treatment of the SYK island \cite{Kruchkov2020}. The first temperature bound is set by the bandwidth $t$. However, the question of the bandwidth for strongly-disorder flakes is not well defined; from from Fig.~\ref{Fig1}b we estimate the upper bounds as $\pm 15$ meV, from which we estimate $t \approx 7.5$ meV; similar upper bounds are imposed by experiment \cite{Zhang2006}. This gives $T_1 = {t^2}/{J} \approx 19 \, \text{K}. $
Around this value, the SYK dynamics crossover to conventional Fermi liquid behavior in a universal manner. The second temperature scale is set by the mesoscopic number of SYK states.  In our case of a large flake ($N\approx35$), $T_2 = {J}/{N} \approx 10 \, \text{K}. $ Below $T_2$, mesoscopic fluctuations  are described by the universal Schwarzian theory of the SYK model (with possible perturbations from $t$)  \cite{Kruchkov2020}. Therefore, we come to conclusion that in the magnetic fields of $16$ T, flake sizes of order 100 nm, the SYK dynamics is most favorable in the regime $20 \, \text{K} \lesssim T_{\text{SYK}}  \ll  300 \, \text{K}$. We hence expect the signatures of SYK model \cite{Gu2020} in this temperature range, and probe the relevant quantum transport, namely conductance and thermopower, in this regime \cite{Kruchkov2020}.  This allows to operate the 100 nm graphene flakes of Figs.~\ref{Fig1}, \ref{Fig2} above the point of the liquid helium temperature ($T_{\text{He}} = 4.2$ K), a relevant experimental benchmark.

\

\textit{Conclusions}. In conclusion, by performing large-scale calculation on up to 300 000  atomic sites, we have demonstrated that the SYK interactions can be controllably engineered and enhanced in the disordered graphene flake in realistic magnetic fields 5-20 T,  when the flake enters the quantum dot regime. The obtained results speak in favor of underlying SYK dynamics in the disordered graphene flakes, establishing realistic experimental conditions in terms of length scales, temperatures, and magnetic fields.  Further theoretical modelling of transport across such disordered graphene flakes is required to interpret the graphene prototypes behavior in experiments~\cite{Anderson2022}.

\

\textit{Acknowledgements}. We thank Philip Kim, Bertrand Halperin, Laurel Anderson for useful discussion.  This work was supported
by the Branco Weiss Society in Science, ETH Zurich,
through the research grant on flat bands, strong interactions and the SYK physics. Computations
have been performed at the Swiss National Supercomputing Centre (CSCS) and the facilities
of Scientific IT and Application Support Center of EPFL.
S.S. were supported by the National Science Foundation under Grant No.~DMR-2002850 and by the Simons Collaboration on Ultra-Quantum Matter, a grant from the Simons Foundation (651440). Y.G. and O.V.Y. acknowledge support from the Swiss National Science Foundation (grant No. 204254).

\bibliography{Refs}

\end{document}